# Experimental probing of the interplay between ferromagnetism and localisation in (Ga,Mn)As

Maciej Sawicki[1,2]*, Daichi Chiba[3,1], Anna Korbecka[4], Yu Nishitani[1], Jacek A. Majewski[4], Fumihiro Matsukura[1,3], Tomasz Dietl[2,4,3], and Hideo Ohno[1,3]

[1]Laboratory for Nanoelectronics and Spintronics, Research Institute of Electrical Communication, Tohoku University, Katahira 2-1-1, Aoba-ku, Sendai 980-8577, Japan.

[2]Institute of Physics, Polish Academy of Sciences, al. Lotników 32/46, PL-02 668 Warszawa, Poland.

[3]Semiconductor Spintronics Project, Exploratory Research for Advanced Technology, Japan Science and Technology Agency, Sanban-cho 5, Chiyoda-ku, Tokyo 102-0075, Japan.

[4]Institute of Theoretical Physics, University of Warsaw, ul. Hoża 69, PL 00 681 Warszawa, Poland

*e-mail: mikes@ifpan.edu.pl

**Abstract**

The question whether the Anderson-Mott localisation enhances or reduces magnetic correlations is central to the physics of magnetic alloys[1]. Particularly intriguing is the case of (Ga,Mn)As and related magnetic semiconductors, for which diverging theoretical scenarios have been proposed[2-9]. Here, by direct magnetisation measurements we demonstrate how magnetism evolves when the density of carriers mediating the spin-spin coupling is diminished by the gate electric field in metal/insulator/semiconductor structures of (Ga,Mn)As. Our findings show that the channel depletion results in a monotonic decrease of the Curie temperature, with no evidence for the maximum expected within the impurity-band models[3,5,8,9]. We find that the transition from the ferromagnetic to the paramagnetic state proceeds via the emergence of a superparamagnetic-like spin arrangement. This implies that carrier localisation leads to a phase separation into ferromagnetic and nonmagnetic regions, which we attribute to critical fluctuations in the local density of states, specific to the Anderson-Mott quantum transition.

Manipulation of magnetism by a gate electric field has been demonstrated in carrier-controlled ferromagnets by studies of the anomalous Hall effect[10-14], resistance[15,16] and splitting of a luminescence line[17]. Such studies provide information on spin polarisation of itinerant carriers. In order to probe directly the effect of carrier localisation on magnetism we have developed superconducting quantum interference device (SQUID) magnetometery sensitive enough to determine quantitatively magnetisation of (Ga,Mn)As consisting the channel of metal-insulator-semiconductor (MIS) structures. A 3.5 thick film of $Ga_{0.93}Mn_{0.07}As$ are grown at 220°C on a 4 nm GaAs/30 nm $Al_{0.8}Ga_{0.2}As$/30 nm GaAs buffer layer structure on a semi-insulating GaAs(001) substrates by molecular beam epitaxy. For magnetic measurements, a series of (Ga,Mn)As-based large parallel-plate capacitors, with an average gate area $A$ of about 10 mm$^2$, with atomic layer deposition grown gate insulator $HfO_2$ of thickness $d$ = 50 mm and the dielectric constant $\kappa \cong 20$ has been prepared. The 3 nm Cr/50 nm Au gate electrode completes the structure. We regard a device as a prospective one if during room temperature current-voltage $I$-$V$ characteristics tests within ±4 V (+/- ~1MV/cm) the capacitor shows no indication of leaking (flat $I$-$V$ 'curve' on 100 pA range indicating the leakage current below $10^{-10}$ A/cm$^2$). Typically one third of thees devices survive multiple mounting on a magnetometer sample holder and low temperature energising up to the gate voltage of 12-15 V (3-4 MV/cm). An employment of indium for contacting (Ga,Mn)As channel and some technical details of the experimental set-up limits reliable magnetometry of these devices to temperatures above 6 K and magnetic fields below a few kOe, respectively.

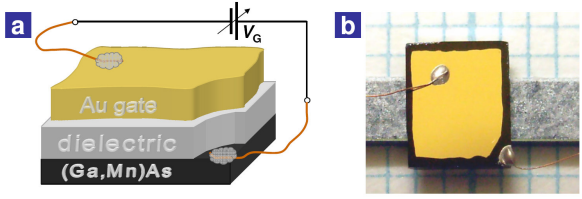

**Figure 1. a,** The arrangement of the main components of devices fabricated for our studies. **b,** A view of one of the devices. The device is attached to the sample holder comprising of a 2 mm wide and 19 cm long silicone strip with an aid of equally long strip of a thin double-sided low tack sticky tape. The connections to the voltage terminals at the far ends of the sample holder are made by a high purity 50 μm copper wire.

Our as-grown material shows the Curie temperature $T_C < 25$ K, a typical value for very thin (Ga,Mn)As films[18], with reduced hole density, bringing the system near the localisation boundary. Because of magnetic anisotropy, the SQUID measurements are performed along various crystallographic directions. The results presented here correspond to the in-plane configuration, as we find that the magnetic moment remains in-plane even at low hole densities for which the easy axis assumes the perpendicular orientation in thick layers.

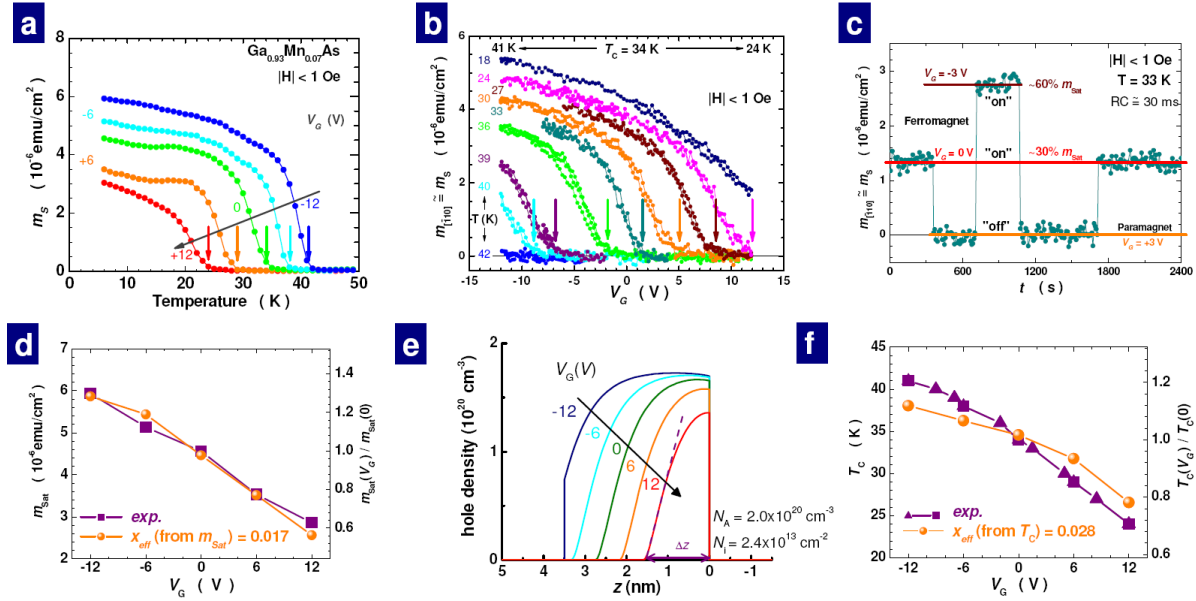

**Figure 2. a,** Experimental temperature dependence of the spontaneous moment $m_S$ for selected values of gate voltage $V_G$. Temperatures at which $m_S$ disappears defines the Curie temperature $T_C$, as marked by arrows. **b,** Isothermal $m_S(V_G)$. Arrows mark the Curie points for each $T$. **c,** Isothermal control of $m_S$ including complete switching 'on' and 'off' of the ferromagnetic state of the gated part of the sample. The temporal resolution of the experiment is limited mainly by the acquisition time of the SQUID, approximately 1.5 s per point. **d,** Experimental and calculated dependence of $m_{Sat}$ on $V_G$. **e,** Profile $p(z)$ of the hole density in the channel for various $V_G$ computed assuming the interfacial donor concentration $N_i = 2.4 \times 10^{13}$ cm$^{-2}$ and the net acceptor concentration $N_A = 2.0 \times 10^{20}$ cm$^{-3}$ in the 3.5 nm-thick channel. The dashed line exemplifies the method used to established $\Delta z$, defining the region populated by holes, required to calculate the saturation magnetic moment $m_{Sat}$. **f,** Experimental dependence of $T_C$ on $V_G$: dark squares (from a) and triangles (from b). Bullets are calculated from Eq. 1 assuming the hole profiles of e.

We obtain $T_C(V_G)$ from the temperature dependence of the spontaneous magnetic moment $m_S(T)$ of the film during cooling the sample in *zero* magnetic field (< 1 Oe) under constant $V_G$ applied at $T > T_C$ (see Fig. 2a) and from measurements of $m_S$ as a function of $V_G$ at various temperatures (see Fig. 2b). From the results presented in Figs. 2a and 2b, we find that processing, presumably unintentional annealing, increases $T_C$ to 34 K at zero gate voltage. Then the magnitude of $T_C$ rises to 41 K under negative bias of $V_G = -12$ V, which increases the hole density, and can conversely be reduced to 24 K under the positive voltage $V_G = +12$ V. The present results, together with those obtained previously[12,13], demonstrate consistently that the value of $T_C$ decreases when lowering the carrier concentration. This contradicts impurity-band models of (Ga,Mn)As ferromagnetism, which predict a ceiling in the $T_C$ values when the Fermi energy passes through a maximum in the density of the impurity-band states[3,5,8,9].

Figure 2c shows the isothermal switching among different magnetic states in the time domain, which provides the absolute scale of the magnetic moment changes reported previously for (In,Mn)As[10,12] and (Ga,Mn)As[13,14]. In Fig. 2a, $m_{Sat} = m_S(T \to 0)$ changes monotonically in the whole range of $V_G$ (see also Fig. 2d), indicating that the channel is partially depleted at $V_G = 0$ since no increase of magnetic response is expected once all the moments are involved in the ferromagnetic phase.

In order to find out to what extend one can describe the data disregarding effects of localisation, we determine the hole density $p$ along the growth direction $z$ by using a Poisson solver under the Thomas-Fermi approximation[19]. Since in (Ga,Mn)As the mean free path is shorter than the channel width, the dimensional quantisation is smeared out, which makes this semi-classical approach valid. We determine $p(z)$ and then $m_{Sat}(V_G)$ and $T_C(V_G)$ in a self consistent way adopting the hole density-of-states effective mass from the 6×6 $kp$ model[2]. Furthermore, since the density of the surface states as well as the degree of compensation are unknown, we treat the concentration of positive interface states $N_i$ and the net concentration $N_A$ of Mn acceptors in the (Ga,Mn)As channel as two adjustable parameters to obtain such $p(z)$ that describes the experimental values of $m_{Sat}(V_G)/m_{Sat}(0)$.

As shown in Fig. 2d, the calculated dependence of the relative magnetic moment $m_{Sat}(V_G)/m_{Sat}(0) = \Delta z(V_G)/\Delta z(0)$, where $\Delta z(V_G)$ is the thickness of the layer populated by the holes (see Fig. 2e), reproduces well the experimental values. This points to a strongly non-uniform hole distribution across the width of the channel for all gate voltages examined here. In addition, the presence of surface depletion layer shows that an efficient steering of the hole density towards enhanced $T_C$ or anisotropy switching at the high-$p$ end[20] requires the reduction of interface states.

In order to determine $T_C(V_G)$ we note that the channel width is narrower than the phase coherence length of the holes[21], which implies a collective two-dimensional behaviour of the Mn spins across the channel. In this case, in terms of the sheet hole density $p_s = \int dz\, p(z)$ and the corresponding thermodynamic density of states at the Fermi level $\rho_s(E_F) = \partial p_s/\partial E_F$, $T_C$ according to the $p$-$d$ Zener model[2,22] can be written in the form,

$$T_C(V_G) = \int dz\, T_C[p(z), x_{eff}] \int dz\, p^2(z)/p_s^2 \qquad (1)$$

where $T_C[p,x_{eff}]$ is the Curie temperature computed earlier[2] for (Ga,Mn)As as a function of the hole concentration and the effective concentration of Mn participating in the ferromagnetic ordering $x_{eff}N_0$, where $N_0$ is the cation concentration. As shown in Fig. 2f, basing on profiles from Fig. 2e, Eq. 1 can explain the dependence of $T_C(V_G)/T_C(0)$ reasonably well. We note also, that while with the fixed values of $N_i$ and $N_A$ we can describe both $m_{Sat}(V_G)/m_{Sat}(0)$ and $T_C(V_G)/T_C(0)$, the fitting could be improved by taking the thickness of (Ga,Mn)As, due to the possible existence of a native oxide layer[16], and/or the concentration of As-antisite donors in the GaAs buffer layer, as additional adjustable parameters.

In (Ga,Mn)As the self compensation proceeds through the creation of Mn interstitial double donors of concentration[23] $x_I N_0 = (xN_0 - N_A)/3$. According to both theoretical[24,25] and experimental[26] findings, they do not couple to the hole spins, so that $x_{eff} = x - x_I$. Indeed, by fitting the absolute values of $T_C(V_G)$ and $m_{Sat}(V_G)$ we obtain reduced values of $x$, $x_{eff}$ (from $T_C$) = 0.028 and $x_{eff}$ (from $m_{Sat}$) = 0.017, as shown in Figs. 2f and 2d, respectively. According to the determined values of $N_A = 2.0 \times 10^{20}$ cm$^{-3}$, $x_I$ is 0.020, which gives $x_{eff} = 0.05$. We see that $x_{eff}$ determined from experiments is smaller than this value, particularly in the case of $m_{Sat}(V_G)$. In search for the missing magnetic moments we have undertaken an elaborated method of magnetisation determination.

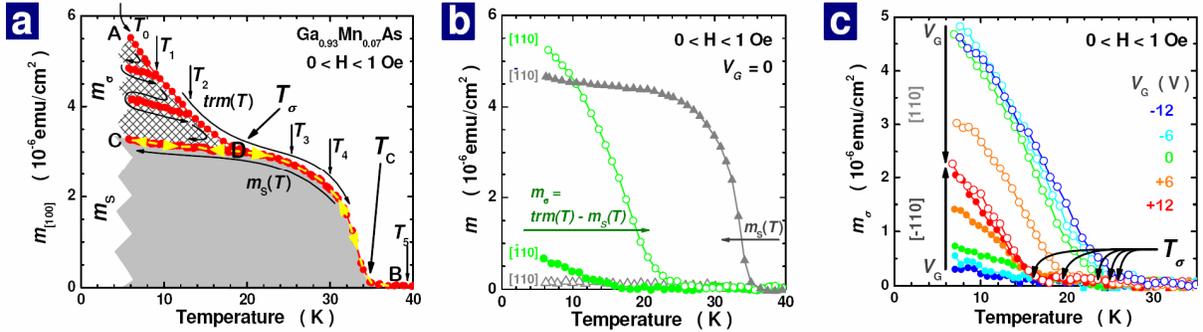

**Figure 3. Experimental evidence for the coexistence of ferromagnetic and superparamagnetic-like regions specified by differing magnetic anisotropy. a,** After field cooling in $H = 1$ kOe, a standard measurement of thermoremanent magnetization [$trm(T)$, red points] along [100] direction starts at the initial temperature $T_0 = 6$ K (point A) after the field quenching. Then instead of warming the system directly to above $T_C = 34$ K ($T_5$, point B) the temperature sweep is interrupted at some intermediate temperatures ($T_1$ to $T_4$) and redirected back to $T_0$ - where the warming is restarted to next $T_i$. After exceeding a characteristic temperature $T_\sigma \approx 19$ K (point D) all the experimental points fall onto the same temperature-reversible line (C-D-B, dashed yellow). The procedure allows then to identify in general: (i) the temperature-reversible component $m_S$ (the grey area under the C-D-B line) and (ii) the history dependent part $m_\sigma$ (crossed hatched triangle A-D-C 'above' the C-D-B line); or in particular their projections. **b,** Temperature dependence [-110] (full points) and [110] (open points) components of $m_S$ (triangles) and $m_\sigma$ (circles), showing that the easy axis for most of $m_\sigma$ is [110], that is perpendicular to the easy axis of $m_S$. **c,** Demonstration that $T_\sigma$ (black arrows) and uniaxial magnetic anisotropy of $m_\sigma$ (difference between open and full points) decreases with the channel depletion, the anisotropy vanishing entirely for $V_G = + 12$ V.

As shown in Fig. 3a, by cooling in the magnetic field we reveal the presence of an increased (with respect to $m_S$) low temperature magnetic remanent signal. This additional contribution becomes visible only below a certain characteristic temperature $T_\sigma < T_C$. While the already discussed spontaneous magnetic moment $m_S$, shows properties expected for standard ferromagnets, the properties of this additional magnetic moment $m_\sigma$ are similar to those exhibited by a collection of magnetic nanoparticles. We hereafter call it a superparamagnetic-like moment and identify $T_\sigma$ as a relevant blocking temperature. Interestingly, as shown in Fig. 3b, the moments $m_\sigma$ (circles) and $m_S$ (triangles) are oriented in the mutually perpendicular directions, along [110] and [-110], respectively.

In order to interpret these observations, we note that according to some of the present authors[8,27], properties of carrier-mediated ferromagnets are strongly influenced by the vicinity of the metal-to-insulator transition (MIT), particularly by the corresponding quantum critical fluctuations in the local density of states. These fluctuations break the spatial continuity of the ferromagnetic order leading to a nanoscale phase separation, a scenario suggested also by Monte-Carlo simulations[13] and x-ray magnetic dichroism[28] in (Ga,Mn)As. This process is seen by inspecting the data in Figs. 2e and 3c for $V_G \leq 0$, where the relative contribution $m_\sigma(T \to 0)$ to $m_{Sat}$ increases on going towards more positive $V_G$. On further depleting of the channel ($V_G > 0$) a reduction in the volume of superparamagnetic regions is expected owing to a shortening of the localisation length. The presence of this effect is corroborated by the low-field studies pointing to a rapid fall of $T_\sigma$ and progressive randomisation of the easy axis orientations, as seen in Fig. 3c.

The model presented here reaches further beyond a very thin films case. For any thickness, the depletion of the near-the-surface or interface will introduce this superparamagnetic-like response in a proportion depending on the thickness of the film. Analogously, if a film of any thickness is brought to the MIT, say, by decreasing $x_{eff}$ and, thus, $p$ and $\rho(E_F)$, its magnetic properties will also be affected by the carrier localisation. However, similarly to other thermodynamic properties, ferromagnetic characteristics do not show up any critical behaviour at the MIT. Instead, within the present scenario, washing out of ferromagnetism by carrier localisation proceeds via nanoscale phase separation resulting in the gradual appearance of superparamagnetic-like properties. This means that in the insulating phase ferromagnetic ordering is maintained only locally but, nevertheless, the apparent $T_C$ is still satisfactorily described by the $p$-$d$ Zener model[8]. A further shortening of the correlation length (which corresponds to a mean localisation length in this case) brings relevant blocking temperatures towards zero, so that *no* magnetic moments are observed without an external magnetic field at non-zero temperatures. Experimentally, a jump of the apparent $T_C$ to zero occurs rather abruptly[15]. In the case of (Ga,Mn)As, where the MIT appears at relatively high values of $x_{eff}$, and $\rho(E_F)$, the reported Curie temperatures for (Ga,Mn)As at the ferromagnetism boundary group around a common value of $T^*_C \cong 15$ K, before jumping to zero. In an agreement with this conjecture, lower than ~15 K values of $T^*_C$ have been reported for, e.g., (Zn,Mn)Te (Ref. 8) and (In,Mn)Sb (Ref. 29), in which the MIT occurs at smaller values of $x_{eff}$ and $\rho(E_F)$, respectively. We note that this reasoning does not rule out a continuous decay of $T^*_C$ with $p$, it rather calls for more elaborate magnetic studies in this region of the magnetic phase diagram.

In summary, we have developed a method to monitor experimentally and to describe theoretically changes in the magnetic moment and ordering temperature induced by an electric field in MIS structures of ferromagnets in which the spin-spin coupling is mediated by carriers. We have found that further progress in magnetisation manipulations by an electric field requires a better control over interface states. Furthermore, our data demonstrate the co-existence of ferromagnetic-like, superparamagnetic-like, and paramagnetic regions. We assign this phase separation to quantum critical fluctuations in the local density of hole states, specific to doped semiconductors in the vicinity of the Anderson-Mott MIT. Within this model, the portion of the spins which contribute to the long-range ferromagnetic order diminishes when electronic disorder increases, which eventually results in the loss of spontaneous magnetisation. This occurs at higher values of disorder than that corresponding to the MIT[30]. According to our results, magnetic anisotropy undergoes significant changes in this region. While the distribution of magnetic ions has been assumed to be uniform, in many systems magnetic ions tend to aggregate. In such a case the magnitude of the apparent Curie temperature is determined rather by a non-random distribution of magnetic ions[27] than by the fluctuations in the local carrier density.

**Acknowledgements** We acknowledge discussion with Y. Norifusa and T. Endoh. The work at Tohoku University was supported by Grant-in-Aids from MEXT/JSPS, the GCOE program, the Research and Development for Next-Generation Information Technology Program (MEXT); M.S, A.K, J.A.M. and T. D. acknowledge financial support from EU (NANOSPIN EC: FP6-IST-015728 and FunDMS Advanced Grant of ERC); A.K. and J.A.M. acknowledge the support of the Polish Ministry of Science and High Education (project no. N2002 02632/0705).